\documentclass[preprint,prl,aps,showpacs]{revtex4}
\begin{document}

\newcommand{\be}{\begin{equation}}
\newcommand{\ee}{\end{equation}}
\newcommand{\ben}{\begin{eqnarray}}
\newcommand{\een}{\end{eqnarray}}
\newcommand{\nn}{\nonumber \\}
\newcommand{\ii}{\'{\i}}
\newcommand{\pp}{\prime}
\newcommand{\expq}{e_q}
\newcommand{\lnq}{\ln_q}
\newcommand{\quno}{q-1}
\newcommand{\qunoinv}{\frac{1}{q-1}}
\newcommand{\tr}{{\mathrm{Tr}}}

\draft

\title{Heisenberg--Fisher thermal uncertainty measure}

\author{F.~Pennini}

\author{A.~Plastino}

 \address{Instituto de F\'{\i}sica La Plata (IFLP)\\
 Universidad Nacional de La Plata (UNLP) and Argentine National
 Research Council (CONICET)\\ C.C.~727, 1900 La Plata, Argentina}


\begin{abstract}
 We establish a connection among i) the so-called Wehrl entropy,
ii) Fisher's information measure $I_\beta$, and iii)  the
canonical ensemble entropy for the one-dimensional quantum
harmonic oscillator~(HO). We show that the contribution of the
excited HO spectrum to the mean thermal energy is given by
  $I_\beta$, while the pertinent canonical partition function is essentially given by another
  Fisher measure: the so-called shift invariant one.
 Our findings should be of interest in view of the fact that it
 has been shown that the Legendre transform structure of
 thermodynamics can be replicated without any change
  if one replaces the Boltzmann--Gibbs--Shannon entropy
   by Fisher's information measure [{\it Physical Review E} {\bf 60}, 48
 (1999)].  New Fisher-related uncertainty relations are
also  advanced, together with a Fisher version of thermodynamics'
third law.

 \end{abstract}
 \maketitle
 \newpage

\section{Introduction}
 We will here  explore some  features of an information-theoretic
 uncertainty measure, the Wehrl entropy~\cite{wehrl}~$I_W$. As
 shown by Lieb \cite{lieb}, $I_W \ge1$, and this bound represents
  a strengthened version of the uncertainty principle. In the case
  of a harmonic oscillator in a thermal state, $I_W$ coincides with
  the logarithmic information measure of Shannon's in the high
  temperature regime. However, it does not vanish at zero
  temperature, thus supplying a nontrivial measure of uncertainty
  due to both thermal and quantum fluctuations~\cite{PRD2753_93}.
 It will be here  shown that intriguing  connections link $I_W$ to Fisher's
 information measure $I$. As far as possible, we will use the notation of
  Anderson and Halliwell~\cite{PRD2753_93}.

 The present endeavor is motivated by the fact that  much
 interesting work has recently been
  devoted to the physical applications of Fisher's information measure (FIM)
 (see, for
  instance,
  \cite{Frieden,roybook, Renyi,FPPS,Incerteza}).  Frieden and Soffer
  \cite{Frieden} have shown that Fisher's information measure provides one
 with a  powerful variational principle,
  the extreme physical information  one,  that yields most of the canonical
 Lagrangians
   of theoretical physics  \cite{Frieden,roybook}.
  Additionally, $I$ has been shown to provide an interesting characterization
  of
  the ``arrow of time'', alternative to the one associated with Boltzmann's
 entropy
  \cite{pla2,pla4}. Thus, unravelling the multiple FIM facets and their
 links to physics should be of general interest. The Legendre
 transform structure of thermodynamics can be
 replicated as well, without any change,
  if one replaces the Boltzmann-Gibbs-Shannon entropy $S$ by Fisher's information measure
 $I$, In particular, i) $I$ possesses the all
  important concavity property
   \cite{FPPS}, and ii)  use of the
  Fisher's measure
 allows for the development of a
  thermodynamics that seems to be able to treat equilibrium and
 none-equilibrium
  situations in a manner entirely
  similar to the conventional one~\cite{FPPS}. Here, the focus of our
  attention will be, following~\cite{PRD2753_93}, the  thermal
  description of harmonic oscillator (HO).

For the convenience of the reader, in Sec. \ref{back} we review
  some fundamental aspects of the  HO canonical-ensemble
   description from a coherent states' viewpoint~\cite{PRD2753_93}. We also
 discuss some ideas related to
  Fisher's information measure,
   the protagonist of the present effort, in Sec. \ref{Fisher}.
  In \ref{law} we  explore its properties with regards to temperature's estimation while, in
   Sec. \ref{uncertainties}, we establish some new results
  concerning  uncertainty relations. Finally, we draw some conclusions in Sec. \ref{conclusions}.

  \section{Background notions}
  \label{back}

  In \cite{PRD2753_93} the authors discuss quantum-mechanical
  phase-space distributions expressed in terms the celebrated
  coherent states $\vert z \rangle$ of the harmonic oscillator,
  eigenstates of the annihilation operator $\hat a$   \cite{klauder,Schnack}

      \ben  \label{z} \hat{H}_o &=& \hbar
    \omega \,[\hat a^{\dagger} \hat a + 1/2];\,\,
     \hat a = i({2\hbar \omega
  m})^{-1/2}\hat p + (m\omega/2\hbar)^{1/2}\hat x \cr
   2 z&=& \left(x/\sigma_x +i p/\sigma_p\right)\cr
  z&=&(m\omega/2\hbar)^{1/2}x + i({2\hbar \omega
  m})^{-1/2}p \cr &\equiv& x'+i p',\,\,\,\,{\rm with}\,\,\,\,
  x'= x/2\sigma_x;\,\,\,\, p'=
  p/2\sigma_p\cr \sigma_x &=&(\hbar/2m\omega)^{1/2};\,\,\,
\sigma_p=(\hbar m \omega/2)^{1/2};\,\,\, \sigma_x \sigma_p
=\hbar/2. \een
  Variances $\sigma$ are evaluated
 for the  HO ground state. Coherent states span Hilbert's space,
 constitute an over-complete basis and obey the completeness rule (use``natural
 variables" $x',\,y'$)
 \cite{klauder}
 \ben \label{klauder} \int \frac{\mathrm{d}^2\, z}{\pi}\,\vert z \rangle\langle z
 \vert &=& \int \frac{\mathrm{d}p\,\mathrm{d}x}{2\pi\hbar}\,\vert
 p,x
\rangle\langle p,x \vert =  1 \cr  \mathrm{d}^2z &=&
\mathrm{d}\,\Re(z)\,\,\mathrm{d}\,\Im(z)\cr
 &=&\frac{\mathrm{d}p\,\mathrm{d}x}{2\hbar}\equiv \mathrm{d}p'\,\mathrm{d}x'. \een
  Variances $\sigma$ are evaluated
 for the  HO ground state.   The Wehrl
  entropy~\cite{wehrl} is defined as
  \be
  I_W=-\int \frac{\mathrm{d}p\, \mathrm{d}x}{2 \pi \hbar} \mu(p,x)\, \ln
  \mu(p,x),\label{i1}\ee
    where $\mu(p,x)=\langle z|\hat \rho|z\rangle$
  is the ``classical'' distribution function associated to the
  density matrix $\hat \rho$ of the system. The function $\mu(p,x)$ is
  normalized in the fashion $ \int (\mathrm{d}p\, \mathrm{d}x/2 \pi
 \hbar)\,
  \mu(p,x)\equiv\int (\mathrm{d}p'\, \mathrm{d}x'/ \pi
 )\,
  \mu(p',x')=1,
  $
  and is often referred to as the Husimi
  distribution~\cite{husimi}.
  It is of particular interest to discuss the equilibrium case, as
  represented by Gibbs' canonical distribution, where the
  ``thermal''
   density matrix is given by  $\hat \rho=Z^{-1}e^{-\beta \hat H}$.
  $Z=Tr(e^{-\beta \hat H})$ is the partition function, and $\beta=1/kT$,
   being $T$ the temperature,
  with $k$ the Boltzmann constant, to be set equal to unity
  hereafter.
  Specializing things for  the HO, with eigenstates    $|n\rangle$
 associated to the eigen-energies $ E_n=\hbar
  \omega\left(n+\frac12\right)$, one
 has $ \langle z|\hat \rho|z \rangle=\frac{1}{Z} \sum_{n} e^{-\beta
\hat H}|\langle z|n\rangle|^2$ with
 $
  |\langle
  z|n\rangle|^2=\frac{|z|^{2n}}{n!}e^{-|z|^2},
  $
so that the distribution $\mu$ reads ~\cite{PRD2753_93}
  \be \mu(p,x)=\langle z|\hat \rho|z \rangle=(1-e^{-\beta \hbar
  \omega})e^{-(1-e^{-\beta \hbar \omega})|z|^2}\label{cstate}, \ee
  and the Wehrl information (\ref{i1}), after integration over
  all phase space, turns out to be \be I_W=1-\ln(1-e^{-\beta \hbar
  \omega}).\label{i0} \ee  Notice that
  $x',\,\,y'$ in (\ref{z}) have been chosen so as to obtain the
  following result. First define (1) $e^2_{\vert z \vert}(\beta,\omega) \equiv e^2_{\vert z
  \vert},$ \newline  (2) $J_2= (1/\pi)\int
  \mathrm{d}p'\mathrm{d}x'\,\mu(p',x')\,\vert z \vert^2$, and \newline (3) $J=(1/\pi)\,\int
  \mathrm{d}p'\mathrm{d}x'\,\mu(p',x')\,\vert z \vert$. Then:
  \be \label{meansqerror}
 e^2_{\vert z \vert} = J_2-J^2
  = \langle \vert z \vert^2 \rangle_\mu
   - \langle \vert z \vert \rangle_\mu^2
   = \int
\frac{\mathrm{d}p'\mathrm{d}x'}{\pi}\,\mu(p',x')\,(p'^2+x'^2) =
     (1-e^{-\beta\hbar\omega})^{-1}.\ee
  We write down now, for future reference,
  well-known  HO-expressions for,
  respectively, the entropy $S$, the mean energy $U$, the specific heat $C_V$,
  and $Z$ ~\cite{pathria1993,Cohen}
   \ben \label{textbook} S&=& \beta \frac{\hbar \omega}{e^{\beta \hbar \omega}-1}-
  \ln{\{1-e^{-\beta \hbar \omega}\}}\cr
  U&=& \hbar\omega\left[\frac12+\frac{1}{e^{\beta\hbar\omega}-1}\right] \cr
  C_V&=& -\beta^2\left(\partial U/\partial
  \beta\right)_V=
  \left[\frac{\hbar\omega\beta}{e^{\beta\hbar\omega}-1}\right]^2e^{\beta\hbar\omega}\cr
  \ln{Z}&=& -\beta\frac{\hbar \omega}{2}- \ln{\{1-e^{-\beta\,\hbar
  \omega}\}}.\een

  \section{Fisher's Information measure}
  \label{Fisher}

  One important information measure is that advanced by  R.~A.
  Fisher in the twenties (a detailed study can be found in
  references~\cite{Frieden,roybook}). Let us consider a system that is specified by a physical
  parameter $\theta$,  while {\bf x} is a stochastic variable $({\bf x}\,\in\,\Re^{N})$
  and
  $f_\theta({\bf x})$ the probability density for ${\bf x}$,
  which depends on the parameter $\theta$.  An observer makes a
  measurement of
   ${\bf x}$ and
  has to best infer $\theta$ from this  measurement,
   calling the
    resulting estimate $\tilde \theta=\tilde \theta({\bf x})$. One
   wonders how well $\theta$ can be determined. Estimation theory~\cite{cramer}
   asserts that the best possible estimator $\tilde
   \theta({\bf x})$, after a very large number of ${\bf x}$-samples
  is examined, suffers a mean-square error $e^2$ from $\theta$ that
  obeys a relationship involving Fisher's $I$, namely, $Ie^2=1$,
  where the Fisher information measure $I$ is of the form
  \be
  I(\theta)=\int \,\mathrm{d}{\bf x}\,f_\theta({\bf
  x})\left\{\frac{\partial \ln f_\theta({\bf x})}{
  \partial \theta}\right\}^2  \label{ifisher}.
  \ee

  This ``best'' estimator is called the {\it efficient} estimator.
  Any other estimator must have a larger mean-square error. The only
  proviso to the above result is that all estimators be unbiased,
  i.e., satisfy $ \langle \tilde \theta({\bf x}) \rangle=\,\theta
  \label{unbias}$.   Thus, Fisher's information measure has a lower bound, in the sense
  that, no matter what parameter of the system  we choose to
  measure, $I$ has to be larger or equal than the inverse of the
  mean-square error associated with  the concomitant   experiment.
  This result, $I\,e^2\,\ge \,1,$ is referred to as the
  Cramer--Rao bound \cite{roybook}.
 A particular $I$-case is of great importance:
 that of translation families~\cite{roybook,Renyi},
  i.e., distribution functions (DF) whose {\it
   form} does not change under $\theta$-displacements. These DF
   are shift-invariant (\`a la Mach, no absolute origin for
  $\theta$), and for them
   Fisher's information measure adopts the somewhat simpler appearance
  \cite{roybook}
   \be\label{shift}
  I(shift\,\,invariant)=\int \,\mathrm{d}{\bf x}\,f({\bf x})\,\left\{\frac{\partial \ln
  f({\bf x})}{
  \partial {\bf x}}\right\}^2.
   \ee

  Fisher's measure is additive  \cite{roybook}: If $x\,\,{\rm
  and}\,\, p$ are independent,  variables, $I(p+x)=I(p)+I(x)$. Notice
  that, for $\theta\equiv\tau=(p,x)$ (a point in pase-space),
 we face  a shift-invariance situation. Since
  in defining $z$ in terms of the variables $x$ and $p$, these are
  scaled by their respective variances (Cf. above the
  definition of $\vert z \rangle$), the Fisher
  measure associated to the probability distribution $\mu(p,x)\equiv\mu(\tau)$ will
  be of the form~\cite{Renyi}
  \be I_{\tau}=\int \frac{\mathrm{d}p\,\mathrm{d}x}{2
  \pi \hbar}\, \mu(p,x) \,{\cal A}, \ee
  with
  \be
  {\cal A}=\sigma_x^2\left[\frac{\partial \ln
  \mu(p,x)}{\partial x}\right]^2 +\sigma_p^2\left[\frac{\partial \ln
  \mu(p,x)}{\partial p}\right]^2.\label{II}
   \ee
  Given the $\mu$-expression  (\ref{cstate}), $I_{\tau}$ becomes \be
  I_{\tau}=1-e^{-\beta \hbar \omega}\label{IF}, \ee
which, in view of (\ref{meansqerror}) immediately yields \be
\label{eficaz} I_{\tau}\,e^2_{\vert z \vert}
(\beta,\omega)=1;\,\,(C-R\,\,bound\,\,reached!). \ee We realize at
this point that the Fisher measure built up with Husimi
distributions is to be best employed to estimate ``phase-space
 position" $\vert z\vert$. Further, {\it efficient estimation is
 possible
for all temperatures}, a rather significant result. Comparison
with Eq.~(\ref{i0}) allows one now to write
 \be
  \label{1res} I_W=1- \ln{[I_{\tau}]}\Rightarrow I_W+\ln{[I_{\tau}]}=1.
  \ee
    Since both $I_W$ and $I_{\tau}$ are
  positive-definite quantities, (\ref{1res}) tells us that {\it they
  are complementary informational quantities}, what one of them
  gains, the other loses.
  Following Anderson {\it et al.} \cite{PRD2753_93} let us
  analyze now the
    high and low temperatures limits. When the temperature goes to zero
  $(\beta\rightarrow \infty)$, $I_{\tau}\approx 1$, its maximum
  possible value, since we know that the ground state will be the
  only one to be populated. If, on the other hand, the temperature
  tends to infinity $(\beta\rightarrow 0)$, then $I_{\tau}\approx
  \beta \hbar \omega$ and tends to zero, because we know before-hand
  that, in the limit, all energy levels will be populated in uniform
  fashion. The uniform distribution is that of maximum ignorance
  \cite{katz}. The range of $I_{\tau}$ is $[0,1]$, that of $I_W$ is
  $[1,\infty]$.  Replacing $I_{\tau}$ into Eq. (\ref{textbook}) we  notice that
 \be I_{\tau}= \frac{e^{-\frac12 \beta \hbar \omega}}{Z},
 \label{probab}
 \ee
 so that it coincides with the canonical ensemble probability
  for finding the system in its ground state.

  \section{Fisher, inverse temperature, and thermodynamics' third law}
  \label{law}

Consider now the general definition
  (\ref{ifisher}) of Fisher's information measure in terms of the
  DF~$\mu(p,x)$ \be I_{\beta}=\int \frac{\mathrm{d}p\,\mathrm{d}x}{2 \pi
 \hbar}\,\mu(p,x)
  \left(\frac{\partial \ln \mu(p,x)}{\partial \beta}\right)^2,
  \label{Ig}   \ee with $\beta \equiv \theta$ is the parameter to be
  estimated. Since
  \be \label{aux} \frac{\partial \ln \mu(p,x)}{\partial \beta}
  =\frac{\hbar\omega}{e^{\beta\hbar\omega}-1}\,[1-(1-e^{-\beta\hbar\omega})\,\vert z \vert^2],\ee
one readily ascertains that i) the $\mu$-mean value of (\ref{aux})
vanishes and ii) \be I_\beta =
\left[\frac{\hbar\omega}{e^{\beta\hbar\omega}-1}\right]^2\,\,\,\,\,\,\left(T=[0,\infty]\rightarrow
  I_{\beta}=[0,\infty]\right)
  \label{Igeneral}, \ee which, in view of (\ref{textbook}),
  entails \be \label{CV}  I_\beta=\frac{ e^{-\beta \hbar \omega}}{\beta^2}\, C_V.\ee

\noindent    Reflection upon the $I_\beta$-range (\ref{Igeneral})
might led one to conclude that it constitutes a {\it Fisher
manifestation of thermodynamics' third law}. Not only Shannon's
measure, but also Fisher's (for the HO, at least) vanishes at zero
temperature. Replacing now (\ref{IF}) and (\ref{Igeneral}) into
  the entropy expression (Cf. (\ref{textbook}))  we immediately arrive at the relation
 \be S= \beta  \sqrt{I_{\beta}} - \ln I_{\tau}.\label{IGshift}
 \ee
 The HO entropy can be
  expressed as the sum of  two terms: one associated with the
  Fisher information $I_{\beta}$ and the other with the Fisher
  information for  translation families $I_{\tau}$
  corresponding to the phase-space variables $(p,\,x)$.
  Using Eq. (\ref{textbook}) we also have
   \be \ln{I_{\tau}}=
  -\beta\frac{\hbar \omega}{2}- \ln{Z}=-[\beta E_{gs}+ \ln{Z}]
\ee
   with $E_{gs}$ denoting the ground state
  energy. Thus,
   \be S= \beta\,\left[\frac{\hbar \omega}{2} + \sqrt{I_\beta}\right]+\ln{Z}, \ee
  which is to be compared to the well known canonical ensemble
  general expression connecting $S$ and the mean energy~$U$~\cite{pathria1993}
    \be S= \ln{Z} + \beta U,\ee we see that   $I_{\beta}$ is related
  to the excited spectrum contribution to $U$ while
  $I_{\tau}$ is to be linked to the partition function.
We will look  now for a new connection between Fisher's measures
$I_{\tau}$ and $I_{\beta}$. From (\ref{Igeneral}) it is possible
to rewrite $I_{\beta}$ in the form \be
I_{\beta}\equiv\left(\frac{\hbar \omega\,e^{-\beta \hbar
\omega}}{1-e^{-\beta \hbar \omega}}\right)^2,\label{Iequiv} \ee
and therefore
 \be
I_{\tau}\,\sqrt{I_{\beta}}= \hbar\,\omega\,e^{-\beta
\hbar\omega}=-\frac{\partial }{\partial
 \beta}(e^{-\beta
\hbar\omega}),
  \label{partifi} \ee i.e, the product on the left-hand-side is
  the $\beta$-derivative of the Boltzmann factor (constant
  energy-wise) at the inverse temperature $\beta$. In other words,
$ I_{\tau}\,\sqrt{I_{\beta}}$ measures the $\beta$-gradient of the
Boltzmann factor.

\section{Uncertainties}
\label{uncertainties} \noindent  We focus attention now on the
actual phase-space variables $x,\,p$ (not on $x',\,p'$), and start
with the obvious results $\langle x \rangle_{\mu}=\langle p
\rangle_{\mu}=0$. We immediately find \be (\Delta_{\mu} x)^2=
\langle x^2 \rangle_{\mu}=\int \frac{\mathrm{d}p\,\mathrm{d}x}{2
\pi\hbar}\,x^2\,\mu(p,x) =\frac{2\, \sigma_x^2}{1-e^{-\beta \hbar
\omega}}.\ee In a similar vein
 \be (\Delta_{\mu} p)^2=\langle p^2
\rangle_{\mu}=\frac{2\, \sigma_p^2}{1-e^{-\beta \hbar
\omega}}, \label{x2} \ee
which entails
 \be
\Delta_{\mu}\equiv \Delta_{\mu} x \,\Delta_{\mu}
p=\frac{\hbar}{1-e^{-\beta \hbar \omega}}= \frac{
\hbar}{I_{\tau}}\,\Rightarrow\,I_{\tau}\Delta_{\mu}=\hbar. \ee We
reconfirm thus the already mentioned fact that phase space
``location" is possible, with Husimi distributions~(HPDF's), only
up to $\hbar$. This is to be compared to the uncertainties
evaluated in a purely quantal fashion, without using~HPDF's. This
is made by recourse to the virial theorem~\cite{pathria1993},
that entails both i) $U=m \omega^2 \langle x^2 \rangle,$ and ii)
$U=\langle p^2 \rangle/ m$ (Cf. Eq. (\ref{textbook})). From these
we easily deduce \be \langle x^2
\rangle=\sigma_x^2\,\frac{e^{\beta \hbar \omega}+1}{e^{\beta \hbar
\omega}-1}\,\,\Rightarrow\,\, \langle x^2
\rangle_{\mu}=\frac{2\,\langle x^2 \rangle}{1+e^{-\beta \hbar
\omega}}, \,\,\,\,{\rm and} \ee
 \be \langle p^2 \rangle=\sigma_p^2\,\frac{e^{\beta \hbar
\omega}+1}{e^{\beta \hbar \omega}-1} \,\,\Rightarrow\,\,  \langle
p^2 \rangle_{\mu}=\frac{2\,\langle p^2 \rangle}{1+e^{-\beta \hbar
\omega}}. \ee  Consequently, \be \Delta x\,\Delta
p=\frac{\hbar}{2}\,\frac{e^{\beta \hbar \omega}+1}{e^{\beta \hbar
\omega}-1}\,\Rightarrow\, \Delta_{\mu}=\frac{2\,\Delta x\,\Delta
p}{1+e^{-\beta \hbar \omega}}. \ee As $\beta \rightarrow \infty$,
$\Delta_\mu$ is twice the minimum quantum value for $\Delta x
\Delta p$, and $\Delta_\mu \rightarrow \hbar$, the ``minimal"
phase-space cell. The quantum and semi-classical results do
coincide at very high temperature, though. Finally,  with the help
of (\ref{Iequiv}) and (\ref{partifi}), one readily can recast
Heisenberg's uncertainty relation as a function of both frequency
and temperature in the fashion
 \be  F(\beta,\omega)=\Delta x\,\,
\Delta
p=\frac{\hbar}{2I_\tau}\,(1+e^{-\beta\hbar\omega})=\frac{1}{2}\left[\frac{\hbar}{I_{\tau}}+\frac{\sqrt{I_\beta}}{\omega}\right],
\label{R7} \ee so that, for $T$ varying in $[0,\infty]$, the range
of possible $ \Delta x\,\, \Delta p$-values is $[\hbar/2,\infty].$
Eq. (\ref{R7}) is a ``Heisenberg--Fisher" thermal uncertainty (TU)
relation (for a discussion of the TU concept see, for instance,
~\cite{mandelbrot,flang}). $F(\beta,\omega)$ grows with $I_\beta$
and diminishes with $I_\tau$.  Note that, {\it for fixed
uncertainty} $F(\beta,\omega)=$ constant, $I_\tau$ and $I_\beta$
play ``parallel"  roles: improving temperature-estimation'
performance (in the sense that $I_\beta$ grows) also enhances that
of phase-space location ($I_\tau$ has to grow as well), and
viceversa.

  \section{Conclusions}
  \label{conclusions}

  We have explored  in this work connections between
   canonical ensemble quantities and two Fisher information measures (FIM), associated to
   the estimation of, respectively, i) phase-space location ($I_\tau$) and
   temperature ($I_\beta$).
Our most important result is, perhaps,  to have shown   that there
exists a ``Fisher-associated" third law of thermodynamics (at
least for the HO). From a pure information-theoretic viewpoint, we
have, in addition, advanced several new results, namely, \newline
{\bf(1)} a
 connection between Wehrl's entropy and $I_\tau$ (Cf. Eq. (\ref{1res})),
 \newline {\bf(2)} an interpretation of $I_\tau$ as the HO's ground state occupation
 probability (Cf. Eq. (\ref{probab})),  \newline {\bf(3)} an interpretation of $I_\beta$
 proportional to the HO's specific heat (Cf. Eq.
  (\ref{CV})),  \newline {\bf(4)} the possibility of
expressing the HO's entropy as a sum of two terms, one for each of
the above FIM realizations (Cf. Eq. (\ref{IGshift})),  \newline
{\bf(5)}  a new form of Heisenberg's uncertainty relations in
Fisher
 terms (Cf. Eq. (\ref{R7})). \newline {\bf (6)} that efficient $\vert z
 \vert$-estimation can be achieved with $I_\tau$ at {\it all}
 temperatures, as the minimum Cramer--Rao value is always reached (Cf. Eq. (\ref{eficaz})).

 These results are, of course, restricted to
the harmonic oscillator. However, this is such an important system
that HO insights usually have a  wide impact, as the HO
constitutes much more than a mere example. Nowadays it is of
particular interest for the dynamics of bosonic or fermionic atoms
contained in magnetic traps~\cite{anderson,davis,bradley} as well
as for any system that exhibits an equidistant level spacing in
the vicinity of the ground state, like nuclei or Luttinger
liquids.

  \end{document}